\documentclass[graybox]{svmult}

\usepackage{mathptmx}       
\usepackage{helvet}         
\usepackage{courier}        
\usepackage{type1cm}        
\usepackage{makeidx}         
\usepackage{graphicx}        
\usepackage{multicol}        
\usepackage[bottom]{footmisc}

\def\be{\begin{equation}} 
\def\ee{\end{equation}}

\def\msun{{\Msun}}

\def\etal{{\it et al.~}}  
  
\def\HH{${\rm {H_2}}\,\,$}

\def\HI{\hbox{H~$\scriptstyle\rm I\ $}} 
\def\HII{\hbox{H~$\scriptstyle\rm II\ $}} 
\def\HeI{\hbox{He~$\scriptstyle\rm I\ $}} 
\def\HeII{\hbox{He~$\scriptstyle\rm II\ $}} 
 
\def\CII{\hbox{C~$\scriptstyle\rm II\ $}} 
\def\CIV{\hbox{C~$\scriptstyle\rm IV\ $}} 
\def\SiIV{\hbox{Si~$\scriptstyle\rm IV\ $}} 
\def\OI{\hbox{O~$\scriptstyle\rm I\ $}}

\def\gsim{\lower.5ex\hbox{\gtsima}} 
\def\lsim{\lower.5ex\hbox{\ltsima}} \def\gtsima{$\; \buildrel > \over 
\sim \;$} \def\ltsima{$\; \buildrel < \over \sim \;$} \def\prosima{$\; 
\buildrel \propto \over \sim \;$} \def\gsim{\lower.5ex\hbox{\gtsima}} 
\def\lsim{\lower.5ex\hbox{\ltsima}} 
\def\simgt{\lower.5ex\hbox{\gtsima}} 
\def\simlt{\lower.5ex\hbox{\ltsima}} 
\def\simpr{\lower.5ex\hbox{\prosima}}   
\def\ie{{\frenchspacing\it i.e. }}  
 \def\gtsima{$\; \buildrel > \over \sim \;$} 
\def\ltsima{$\; \buildrel < \over \sim \;$} 
\def\gsim{\lower.5ex\hbox{\gtsima}} 
\def\lsim{\lower.5ex\hbox{\ltsima}} 
\def\simgt{\lower.5ex\hbox{\gtsima}} 
\def\simlt{\lower.5ex\hbox{\ltsima}} 
\def\simpr{\lower.5ex\hbox{\prosima}}

\def\zcr{Z_{\rm cr}}

\def\Lya{Ly$\alpha$~}

\def\msun{\,{\rm \Msun}} 
\def\ie{{\frenchspacing\it i.e., }} 
 
\def\E3{{\cal E}_{\rm g}^{III}}

\def\Msun{\rm M_\odot}

\def\zvir{z_{vir}}

\makeindex             

\begin{document}

\title*{Metal Enrichment in the Reionization Epoch}
\author{Andrea Ferrara}
\institute{Scuola Normale Superiore, Piazza dei Cavalieri 7, 56126 Pisa, Italy, \email{andrea.ferrara@sns.it}}
\maketitle

\abstract{Heavy elements are important constituents of the Universe predominantly produced by massive stars during their evolution. Massive stars are also considered to be primary sources of the ionizing radiation required to power cosmic reionization. Therefore, an intimate link between metal and ionizing photon production must exist. In this chapter, I elaborate on this concept, outlining the basic ingredients necessary to model metal enrichment and interpret experimental data. I conclude with a brief overview of recent theoretical and observational progresses in the field.}

\section{Introduction} 
The presence of elements heavier than helium (``metals'') is of fundamental importance for a large number of astrophysical processes occurring in planet, star and galaxy formation; it also affects cosmic structure formation and evolution in several ways. Even a small amount of heavy elements can dramatically alter the chemistry of the gas, opening the path to complex molecules. Metals might enhance the ability of the gas to radiate away its thermal energy, thus favoring the formation of gravitationally bound objects; they can also condensate in a solid phase (dust grains), partly or totally blocking radiation from luminous sources. Finally, they represent useful tracers of energy deposition by stars and probe the physical properties of the environment by absorption or emission lines.  Last, but certainly not least, life -- as we know it on Earth -- is tightly related to the presence of at least some of the heavy elements. 

In this review I will concentrate on the connection between early metal enrichment and cosmic reionization. As we will see these two processes are intimately connected and their joint study might turn out to be fundamental in understanding the overall evolution of the Universe during the first  billion years after the Big Bang, an epoch corresponding to redshifts $z \ge 6$. 

\section{Big Bang Nucleosynthesis}  
The Earth is made of a large number of different chemical elements, each of them with a different atomic weight, $A$. The same elements are found essentially in every system (gas, planets, stars, galaxies, clusters of galaxies) we have explored so far, independently on their distance, size or age. We usually define elements with $A > 4$ as ``heavy elements'' to distinguish them from the lighter H, He species (keep in mind that nuclei with $A=5$ and $A=8$ are unstable).  

There is an important physical reason to make this distinction. In fact, the abundance of all heavy elements shows abundance variations spanning almost 3 orders of magnitude in different astrophysical environments where we have been able to measure them. In contrast, in the same environments $^4$He abundance with respect to H has a remarkably constant value. Thus the origin of this element has been speculated to result from a cosmological, rather than stellar, process.  The primordial $^4$He abundance by mass has now accurately measured to be $Y=0.2485 \pm 0.0002$ after the exquisitely precise value of the baryon-to-photon ratio of $\eta = (6.08 \pm 0.07) \times 10 ^{-10}$ measured by PLANCK (Planck Collaboration 2013). 
The hot Big Bang model nicely gives a simple explanation for this evidence: He must have formed formed soon (about 3 minutes) after the Big Bang, when the entire Universe was a nuclear reactor, a phase known as the Big-Bang Nucleosynthesis (BBN). This explains why its abundance has a universal value. If thermonuclear reactions are responsible for He production, were they also able to form other elements? 

The next elements are $^{6,7}$Li, $^9$Be and $^{10,11}$B, before getting to carbon.  These species are very fragile and they are not produced in the normal course of stellar nucleosynthesis; actually they are destroyed in stellar interiors. This characteristic is reflected in the low abundance of these simple species. As expansion cooled down matter to temperatures at which nuclear reactions effectively stopped, primordial thermonuclear fusion become unable to proceed efficiently beyond $^{7}$Li, whose primordial abundance remains nevertheless extremely small ($\approx 10^{-9}$). BBN is ineffective in generating $^{6}$Li, $^9$Be and $^{10,11}$B. 

Carbon is produced along inefficient paths involving intermediate mass elements, in particular $^{11}$B rather than the usual 3$\alpha$ reaction in stars. Very small traces of N and O are then produced by radiative capture upon $^{12}$C. Iocco et al. (2007) re-analyzed BBN by adding to the standard code 4 nuclides and more than 100 reactions. Their conclusions confirm those from previous, less detailed studies in obtaining negligible abundances of ($^{12}$C, $^{14}$N, $^{16}$O)/H =$ (4.4 \times 10^{-16}, 2.6 \times 10^{-17}, 1.8 \times 10^{-20}$).  As we will see later, these predictions are relevant for determining the physical conditions of the cosmic gas out of which the first stars form. The gap from  $^4$He to  C, N and O had to await for nuclear reactions in stellar interiors to be bridged, along with the formation of all other heavy elements. 

\section{The beginning of the metal age}  
The next question is therefore, when did the first stars form and start to produce metals? An answer to this question can be in principle obtained from the theory of the growth of gaussian density perturbations. A dark matter halo of total mass $M_h$, representing a $\nu\sigma$ fluctuation of the density field, collapses at a redshift implicitly given by
\begin{equation}
\label{nu}
\nu = \frac{\delta_c(z)}{\sigma(M)} =  \frac{\delta_c}{D(z)\sigma(M_h)},
\end{equation}
where $\delta_c=1.689$, $D(z)$ is the linear growth factor of perturbations, and $\sigma$ is the present-day mass variance of the linear density field. Usually a fiducial value of $\nu=3$ is used, as higher $\sigma$ peaks would collapse even earlier, but they would become too rare to be statistically significant. In general, $D(z)$ must be computed numerically in models with a non-vanishing cosmological constant as a solution to the growing mode amplitude of isentropic perturbations in a pressureless fluid
\begin{equation}
\label{Dz}
D(z) = H(z) \int_z^\infty \frac{(1+z')}{E^3(z')} dz';
\end{equation}
where
\begin{equation}
E(z) =  [\Omega_\Lambda + \Omega_m(1+z)^3]^{1/2}. 
\end{equation}
For $z\gg 1$ an excellent approximation (precise within 1\%) is given by
\begin{equation}
\label{Dz1}
D(z) \approx \Omega_m^{-1/4} (1+z)^{-1}.
\end{equation}.
The variance of the density field is related to the amplitude of the linear power spectrum $P(k) \propto k^n$ by
\begin{equation}
\label{sig}
\sigma^2(R) = \int_0^\infty \Delta^2(k) {\tilde W_G(kR)} \frac{dk}{k}  .
\end{equation}
where $2\pi^2\Delta(k)^2 = k^3 P(k)$ and ${\tilde W_G(kR)}$ is a Fourier-transformed 
top-hat window function. In the subgalactic mass range, an appropriate value for 
the (effective) power spectrum index is $-3 < n_{eff} \le -1$. It is then easy to show that 
\begin{equation}
\label{sig1}
\sigma(R) = \sigma_8 \left(\frac{M_h}{M_8}\right)^{-\beta}
\end{equation}
where $M_8 = {\bar \rho}(t_0) V(R_8) = 3.88\times 10^{14} \msun$ is the total mass 
contained in a present-day sphere of radius $8h^{-1}$ Mpc and $\beta =(n_{eff}+3)/6$.
By combining eqs. (\ref{nu}), (\ref{Dz}), (\ref{sig1})  we obtain the virialization 
redshift of a halo of mass $M_h$,
\begin{equation}
\label{virzm}
(1+z) = \frac{\nu \sigma_8}{\delta_c \Omega_m^{1/4}} \left(\frac{M_h}{M_8}\right)^{-\beta},
\end{equation}
To obtain a quantitative estimate, we fix $n_{eff}=-2.2$ and insert  the relevant cosmological parameters\footnote{Throughout the paper,  we assume a flat Universe with cosmological parameters  given by the   PLANCK13 (Planck Collaboration 2014) best-fit values:  $\Omega_m=0.3175$, $\Omega_{\Lambda} = 1 - \Omega_m=0.6825$,  $\Omega_b h^2 = 0.022068$, and $h=0.6711$.  The  parameters defining  the linear dark  matter power spectrum are $\sigma_8=0.8344$, $  n_s=0.9624$. } in the previous equation. We find that a $M_h=7\times 10^5 \Msun$ halo representing a 3$\sigma$ (5$\sigma$) fluctuation collapses at $\zvir \approx 28$ ($\zvir \approx 47$). More refined calculations confirm this very early appearance of the first stars on the cosmic stage: Naoz et al. 2006 (see also Gao et al. 2007) find that the median
redshift at which there is a 50\% chance of forming the first star is $z=65.8$, which corresponds to a cosmic age of 31 Myr, less than a
0.25\% of the current cosmic age of 13.7 Gyr. Hence, just a few tens of Myr after the Big Bang the Universe starts to become enriched and enters the \textit{metal age}. 

The bulk of metals is produced by massive stars. Among these, for those more massive than about 8 $\Msun$ the final point of the evolution is a supernova explosion through which the nucleosynthetic products are ejected into the surrounding gas. Also, note that the relatively short duration of the reionization epoch (about 1 Gyr, corresponding to $z\approx 6$) compared to the longer evolutionary timescales of intermediate and low-mass stars implies that metal enrichment is controlled by core-collapse supernovae only. At the same time, hydrogen (and in some cases also helium) ionizing photons with energy $h\nu > 13.6$ eV $\equiv $ 1 Ryd are also predominantly produced by the same massive stars, due to their high effective temperatures, and hence hard spectra. The combination of these two evidences makes immediately clear the main point of this review: if metal enrichment and cosmic reionization are produced by the same sources (massive stars) their evolution is intimately linked. 

Obviously, alternatives to this scenario exist, although they do not appear to be favored at present. The first is to consider accreting objects (i.e. black holes powering quasar luminosity) as the reionization sources, thus breaking the link with metal production (for a recent review see Giallongo \etal 2012). In addition, one might consider more exotic scenarios in which decaying/annihilating dark matter particles (Mapelli, Ferrara \& Pierpaoli 2006; Furlanetto \etal 2006; Evoli \etal 2014) might provide the required ionizing photons. Although it is impossible to firmly exclude these options, they appear to fall considerably short of providing the necessary ionizing power to account for the Gunn-Peterson test and CMB polarization data, along with additional constraints coming from the unresolved X-ray background intensity (Salvaterra, Haardt \& Ferrara 2005; McQuinn 2012). Thus, in what follows we will consider stars in galaxies as the only reionizing sources.    

\subsection{Forging metals}
By definition, the first stars contain virtually zero metals (as predicted by BBN, see above; these are referred to as Pop III stars), but later stellar generations (Pop II/I stars, \ie stars with initial metallicity $Z \ge \zcr= 10^{-5} Z_{\odot}$) inherit heavy elements produced by earlier stars and mixed with the gas out of which they formed. The metal content of a system of total gaseous mass $M_g$ is usually quantified by its \textit{metallicity}, defined as
\begin{equation}
Z = \sum_\mathit{i>He} \frac{M_i}{M_g}
\end{equation}
where $i$ runs on all chemical elements heavier than He. It is also common to define the metallicity with respect to the solar value, recently revised downwards, of $Z_\odot = 0.0122$ (Asplund \etal 2005). Alternatively, we can express the abundance of a given element (a commonly used metallicity proxy is iron) relative to the Sun as 
\begin{equation}
\mathrm{\left[Fe/H\right]} = \log_{10}\left(\frac{n_{Fe}}{n_H}\right) - \log_{10}\left(\frac{n_{Fe}}{n_H}\right)_\odot.
\end{equation}
The amount of metals produced by a given star depends on its mass and initial metal content, and supernova explosion energy. In turn, the relative number of massive stars depends on the stellar Initial Mass Function (IMF), i.e. the fractional number of stars as a function of their mass, which is usually approximated as a power-law
\begin{equation}
\phi(M) \equiv \frac{dN}{dM} \propto M^{-x}
\end{equation}
normalized so that,
\be
\int_{M_l}^{M_u} dM  \phi(M)=1.  
\ee 
The present-day Pop II IMF is characterized by the ``Salpeter'' slope $x = 1.35$ (Salpeter 1955). However, it is likely that the IMF of the first Pop III stars was different. This is expected mainly as a result of the different physical conditions and processes regulating the formation of these pristine objects. For example, the lack of cooling (dominated by \HH molecular line emission in a metal-free gas) provided by heavy elements and dust forced the gas in the protostellar environment to higher temperatures. In turn, these would favor a larger (up to $0.1-1 \Msun$ yr$^{-1}$) accretion rate due to the $T^{3/2}$ dependence of this quantity. Hence, more massive stars, up to $\approx 300 \Msun$ could form and the IMF could become flatter or \textit{top-heavy}. It is not yet possible to predict the Pop III IMF in any detail, but simulations (Greif \etal 2012; Hosokawa \etal 2012; Greif 2014)  are rapidly improving and will soon hopefully be able to provide some more insights. 

This limitation does not greatly impact our understanding of cosmic metal enrichment, as the transition from Pop III stars to a more standard Pop II generation occurs at a critical metallicity of $\zcr \approx 10^{-5 \pm 1} Z_{\odot}$ (Schneider \etal 2002, 2006). Let us consider the case of a massive Pop III supernova ejecting its metals in the surrounding gas. In the range $140 < M/\Msun < 260$ Pop III stars encounter a strong instability due to pair-production which causes them to explode leaving no remnant (Heger \& Woosley 2002); these are therefore called pair-instability supernovae (PISN). The explosion will drive a shock wave collecting the surrounding gas into a shell which could eventually fragment  and form new stars (Salvaterra \etal 2004). 

An estimate of the mean metallicity of the swept-up gas, can be obtained from simple dimensional arguments. The amount of metals produced by a PISN is  $M_Z \approx 2 E_{51} \Msun$, where $E_{51}= E/10^{51}$ erg is the total supernova energy, of which a fraction $f_w$ goes into kinetic energy. These heavy elements will be dispersed inside a volume $V \approx  f_w E/\rho c_s^2$, where $\rho$ and $c_s$ are the density and sound speed of the surrounding gas; hence, the corresponding shell mass is $M_s= \rho V = f_w E/c_s^2$. This estimate must be seen as an upper limit to the shell mass, as the previous equation describes a purely adiabatic expansion, wheres it is likely that radiative losses in practice decelerate the expansion of the shock wave. It follows that $ \langle Z \rangle = M_Z/M_s > 3.2 \times 10^{-6} (c_{s,kms}^2/f_w) Z_\odot$, in which the sound speed is in units of km s$^{-1}$. By inserting the fiducial numerical values $c_{s,kms}=10$ and $f_w=0.1$ we find that the shell gas will be enriched to a minimum metallicity of $3.2 \times 10^{-3} Z_\odot \gg \zcr $. 

Although this estimate is uncertain as it assumes a perfect mixing of the metals with the shell gas, the result seems solid enough to allow us to conclude that a single Pop III star is sufficient to prevent further Pop III star formation in a surrounding gas mass $\approx 5 \times 10^4 \Msun$ for a typical PISN with $E_{51}=10$. Thus, if on average a PISN converts 45\%  of its mass into metals, to enrich all cosmic baryons at a mean metallicity equal to $\zcr$, it will only take about 15 PISN cMpc$^{-3}$. This is equivalent to require that we put at least one PISN in each $10^8 \Msun$ halo at $z=15$ to fully quench further Pop III formation (this estimate also neglects additional metal production by coeval Pop II stars). In practice, the process is more gradual as it is difficult to evenly spread the metals over large volumes. However, detailed numerical simulations (Tornatore \etal 2007; Oppenheimer \etal 2009; Xu \etal 2013, Pallottini \etal 2014a) confirm out basic conclusion that metal enrichment by Pop III stars must be negligible, at least on cosmological scales. Therefore, from now on we turn our attention to Pop II stars.

For Pop II stars the standard choice is the canonical Salpeter IMF with lower (upper) mass limit $M_l=0.1\msun$ ($M_u=100\msun$), 
The IMF-averaged Pop II supernova (SNII) yield of a given heavy element $i$ (in solar masses) is then 
\be
y_i \equiv \frac{\int_{\small M_{SNII}}^{\small M_{BH}} dM \phi(M) M_i }
{\int_{\small M_{SNII}}^{\small M_{BH}}dM \phi(M)}
\label{eq:yII}
\ee 
where $M_i$ is the total mass of element $i$ ejected by a progenitor with mass $M$.  The mass range of SNII progenitors is usually assumed to be $(8 - 100) \msun$.  However, above $M_{BH}=50 \pm 10 \msun$ stars form black holes without ejecting heavy elements into the surrounding medium (Tsujimoto \etal 1995), and from $(8-11) \msun$ the pre-supernova evolution of stars is uncertain, resulting in tabulated yields only between $12\msun$ and $40 \msun$ (Woosley \& Weaver 1995). SNII yields may depend on the initial metallicity of the progenitor star. Here we assume for simplicity that SNII progenitors form out of a gas with $Z\ge 10^{-2} Z_{\odot}$.  This choice is motivated by the typical levels of pre-enrichment from PopIII stars (see above). Moreover, the predicted SNII yields with initial metallicity in the range $10^{-4} Z_{\odot} \leq Z \leq 10^{-2} Z_{\odot}$ are largely independent of the initial metallicity of the star (Woosley \& Weaver 1995); above this range variations up to $\approx 10$\% are found as seen in Table 1, where  different combinations of progenitor models relevant for the present analysis are reported.. Finally, we assume that each SNII releases $E=1.2 \times 10^{51}$~erg, independent of the progenitor mass (Woosley \& Weaver 1995).
\begin{table*}
\caption{\footnotesize The IMF-averaged metal yields (in $\msun$) for two models of SNII and SNIa. In the last column $\nu$ is the number of SNII per unit stellar mass formed. }
\begin{center}
\begin{tabular}{@{}ccccccccccc@{}}\hline 
& & & & & & & &\\ [1pt]
SN Model & $M_{SNII} $ & $M_{BH}$ & $Z $ &  $y_{\rm O}$   &  $y_{\rm Si}$ & $y_{\rm S}$  & $y_{\rm Fe}$  &    $y_{\rm Z}$ & $\nu$\\
& $\left[M_\odot\right]$ & $\left[M_\odot\right] $ & $\left[Z_\odot\right]$  & & & & &  & $\left[M_\odot^{-1}\right] $\\ [3pt]
\hline \hline
& & & & & & & &\\ [1pt]
SNII-A   &   12       &   40     &  $10^{-2}$    &   1.430    & 0.133     & 0.064      & 0.136     &     2.03  & 0.00343   \\
SNII-B   &   12       &   40     &    1          &   1.560    & 0.165     & 0.078      & 0.115     &     2.23    & 0.00343 \\
SNIa     &  &  &  &   0.148   & 0.158     & 0.086      & 0.744     &     1.23     \\ \hline
\end{tabular}
\end{center}
\end{table*}
Table 1 shows the IMF-averaged yields for some elements
relevant to our investigation as well as the total mass of metals
released in different SNII models. In the same Table, we also show the
corresponding yields for Type Ia SNe (SNIa), whose values are
mass-independent and have been taken from Gibson, Loewenstein, \&
Mushotzky (1997).

\section{Ionizing photons}
The arguments presented in the previous Sections, suggest that PopIII stars could have been much more massive than stars formed today, with a tentative mass range $30 < M/M_\odot < 1000$. For such a massive star with the concomitant high interior temperatures, the only effective source of opacity is electron scattering.  In the absence of metals and, in particular, of the catalysts necessary for the operation of the CNO cycle, nuclear burning proceeds in a nonstandard way. At first, hydrogen burning can only occur via the inefficient p-p chain. To provide the necessary luminosity, the star has to reach very high central temperatures ($T > 10^8$ K). These temperatures are high enough for the simultaneous occurrence of helium burning via the triple-$\alpha$ process. After a brief initial period of triple-$\alpha$ burning, a trace amount of heavy elements has been formed. Subsequently, the star follows the CNO cycle. The resulting structure consists of a convective core, containing about 90\% of the mass, and a thin radiative envelope. As a result of the high mass and temperature, the stars are dominated by radiation pressure and have luminosities close to the Eddington limit $L_{edd} = 10^{38}(M/M_\odot)$~erg s$^{-1}$.

Metal-free stars with mass above 300 $M_\odot$ resemble a blackbody with an effective temperature of $\approx 10^5$ K, with a production rate of ionizing radiation per stellar mass larger by 1 order of magnitude for H and \HeI and by 2 orders of magnitude for \HeII than the emission from Pop II stars. In the less extreme case of metal-free stars with masses $< 100 M_\odot$, the H-ionizing photon production takes twice as long as that of Pop II to decline to 1/10 of its peak value. Nevertheless, due to the red-ward stellar evolution and short lifetimes of the most massive stars, the hardness of the ionizing spectrum decreases rapidly, leading to the disappearance of the characteristic \HeII recombination lines after about 3 Myr in instantaneous burst models. 

For these reasons, nebular continuum emission cannot be neglected for metal-poor stars with strong ionizing fluxes, as it increases significantly the total continuum flux at wavelengths red-ward of Ly$\alpha$ and leads in turn to reduced emission line equivalent widths. Nebular emission has been included in a more complete and extended study by Schaerer (2002, 2003), who presents realistic models for massive Pop III stars and stellar populations based on non-LTE model atmospheres, recent stellar evolution tracks and up-to-date evolutionary synthesis models, including also different IMFs. Ciardi \& Ferrara (2005) (Table 1) gives a summary of the  emission properties of Pop III stars. The numbers have been derived by integrating the ionizing photon rate in the absence of stellar winds over three different IMFs, i.e. Salpeter, Larson and Gaussian.

For Pop II stars, the ones we are mostly concerned here for the reasons give in the previous Section, the calculation of the number of ionizing photons per baryon into stars is much more solid and straightforward. The ionizing photon production rate, $Q(t)$, by a stellar cluster can be computed exactly from population synthesis models: we use here, as an example, \texttt{Starburst99}\footnote{http://www.stsci.edu/science/starburst99/}. The time dependence of the production rate of Lyman continuum photons under these conditions is, 
\be 
\dot{\cal N}_\gamma(t) = \frac{{\cal N}_0}{1+(t/t_0)^4} ,
\label{eq8}
\ee 
with (${\cal N}_0, t_0$) = ($10^{47} \rm s^{-1} \msun^{-1}, 10^{6.6}$ yr). For consistency with the metal yield calculation above we have used the same Salpeter IMF with lower (upper) mass limit $M_l=0.1\msun$ ($M_u=100\msun$). Eq. (\ref{eq8}) illustrates the important point that after $\sim 4$ Myr, the production rate of ionizing photons rapidly drops as a result of the death of short-lived massive stars. From the same equation it is easy to show that the number of ionizing photons emitted per baryon incorporated into stars is $N_\gamma \approx 0.5 {\cal N}_0 t_0 m_p/\msun =5\times 10^3$.

\section{The reionization link} 
We are now ready to study the connection between metal enrichment and cosmic reionization. As we have pointed out already, it should be clear why the two processes are intimately connected. Once the first stars form, they start to produce ionizing photons that carve \HII (and possibly He-ionized) bubbles in the surrounding intergalactic medium, thereby starting the reionization process. To achieve a complete reionization it is necessary to provide at least one $> 13.6$ eV photon to each hydrogen atom in the Universe. However, this turns out to be just a lower bound as one has to account for the fact that protons and electrons tend to recombine. Thus in general, it is necessary to provide $\kappa$ ionizing photons to each H atom, with $\kappa >1$. 

As massive stars complete their life-cycle and have radiated all their ionizing power, they either collapse into a black hole or explode as a core-collapse supernova (we are not considering anymore the case of massive Pop III stars ending in a PISN). In the previous Sections we have derived in detail the amount of metals produced per baryon going into by these stars.

It is then tempting to combine these two types of information and ask the question: what is the expected mean metallicity of the IGM once reionization is completed? The mass of metals produced by a Pop II stellar population of total mass $M_*$ with the Salpeter IMF assumed here is 
\be 
M_Z = \nu y_Z M_* = \mu m_p \nu y_Z {\cal N}_{b,*} = \mu m_p \nu y_Z \frac{{\cal N}_{\gamma,*}}{N_\gamma}     
\ee 
where  $\mu =1.22$ is the mean molecular weight of a neutral mixture of H and He with a He mass fraction $Y=0.2477$,  $\nu$ is the number of SNII per unit stellar mass formed, ${\cal N}_{b,*}$ is the number of baryons into stars and ${\cal N}_{\gamma,*} = N_\gamma {\cal N}_{b,*}$ is the number of H-ionizing photons produced. We then require that $\kappa(z_r)$ photons must be provided to each H atom by the end of reionization, and find
\be 
M_Z = \mu m_p \nu y_Z \frac{\kappa(z_r) {\cal N}_H}{(1-Y) N_\gamma}     
\ee 
By finally recalling that $Z = M_Z / M_H = M_Z/m_p  {\cal N}_H$ we obtain the following expression for the mean gas metallicity
\be 
\langle Z \rangle = \frac{\mu \nu y_Z \kappa(z_r)}{(1-Y) N_\gamma} = 1.86 \times 10^{-4}  \kappa(z_r) \,\, Z_\odot;    
\label{key}  
\ee 
for the numerical estimates we have used the data from model SNII-A in Table 1. Note that the previous result is rather robust with respect to the IMF assumption made. It can be shown that it would not change substantially if we had instead considered Pop III stars: this is due to the fact that the ratio of metal-to-photon production ($\propto \nu y_Z/N_\gamma$) is close to constant between the two populations. 

Thus, the conclusion is that by the end of reionization (for which we have not yet established the redshift $z_r$) the IGM would be enriched, on average, to a rather substantial metallicity of $\approx 10^{-4}  \kappa ~ Z_\odot$. Note that this values exceeds $\zcr$. Although this does not exclude that under-enriched pockets exist in which Pop III stars can still form, their star formation rate is very likely to be negligible in the post-reionization epoch.

We can push the previous analysis a step further and try to evaluate $\kappa(z_r)$. The hydrogen recombination time can be written as
\be
t_{rec}^{-1} = n_e \alpha_H^{(B)} C_H
\ee
where $n_e$ is the electron density in the IGM and $\alpha_H^{(B)}(T) =  2.59 \times 10^{-13} T^{−0.845}$ cm$^3$ s$^{-1}$, scaled to an IGM temperature of $T = 10^4 T_4$ K. For typical ionization histories and photoelectric heating rates, the diffuse photoionized filaments of hydrogen have temperatures ranging from 5000 K to 20,000 K (Smith et al. 2011; Becker et al. 2011).   The quantity $C_H$ is the so-called hydrogen \textit{clumping factor} and describes the enhancement of the recombination rate due to density inhomogeneities: $C_H \equiv \langle n_e^2 \rangle / \langle n_e \rangle^2$. Some uncertainty exists on the valus of  $C_H$ in the IGM; in addition, it also mildly depends on the reionization history of the gas and on the cosmic UV background intensity (Pawlik \etal 2009; Shull \etal 2012). The latter authors find that $C_H \approx 3$, a consistent downward revision of previous studies that could not properly catch the effect of photoheating raising the cosmological Jeans mass suppression of small-scale density fluctuations. In addition, a higher temperature results in a decreased recombination rate and forces the filaments to expand, hence further reducing $C_H$. In addition, the previous simple analysis neglects the possible correlation between star-forming galaxies and recombining systems which Sobacchi \& Mesinger (2013, 2014) suggest to be important.

As $\kappa$ is defined as the number of recombinations a H-atom undergoes before the end of reionization, this can be then approximated as the ratio $t_H(z_r)/t_{rec}$. As the Hubble time can be written as  $t_H(z_r)^{-1} \approx H_0 \Omega_m^{1/2} (1+z_r)^{3/2}$, it follows that
\be
\kappa(z_r) = 2.96 \left(\frac{1+z_r}{7}\right)^{3/2}  
\ee
or
\be 
\langle Z \rangle = 5.5 \times 10^{-4}  \left(\frac{1+z_r}{7}\right)^{3/2}  \,\, Z_\odot.     
\ee 
Thus an earlier (later) reionization implies in a larger (lower) IGM metallicity. Although this might seem initially counterintuitive, it can be understood by recalling that the mean cosmic density, and therefore the recombination rate, increases with redshift. Thus, more photons per H-atom must be provided, 
which in turn increases the required Lyman continuum photon budget and associated heavy element production.   

\section{Metal dispersal: winds}  
In the previous Section we have seen that by the end of the reionization epoch the IGM metallicity is on average at a level close to the most metal-poor stars in the Milky Way (Salvadori et al. 2007; Salvadori \& Ferrara 2009). However, the distribution of metals is far from homogeneous, and in fact it shows a very patchy structure made of overlapping ``metal bubbles'', in a fashion closely resembling the ``ionized bubbles'' produced by ionizing sources. We will elaborate on this point later on, when presenting the results of specific numerical studies in which the patchy metal enrichment is revealed in its full complexity. 

Metals are dispersed from their formation sites (stars in galaxies) into the IGM by powerful winds driven by the supernova energy deposition. The application of the previous theory to cosmological explosions has to cope with the fact that the background medium is expanding. The simplest explosion models which describes such a situation is a three-component model (Madau, Ferrara \&  Rees 2001) made of (i) a dense, cool spherical shell of outer radius $R_s$ and thickness $\delta R_s$, containing a fraction $(1-f_m)$ of the total baryonic mass enclosed; (ii) a uniform intergalactic medium of density $\rho_m=\rho_b + \rho_d$, including the contribution of baryonic and dark matter, respectively; (iii) a hot, isothermal plasma of pressure $p_b$ and temperature $T_b$ inside the shell. The shell is essentially driven by the thermal pressure of the interior gas, which has to overcome the inertia of the swept up material and gravity force. 
If one assumes that the shell sweeps up almost all the IGM gas ahead, then its mass can be written as $M_s(t) = (4/3)\pi R_s^3 (1-f_m) \rho_b$, with $f_m \ll 1$. 
One can then write the mass, momentum and energy conservation for the shell motion in an expanding universe, for which $\dot\rho/\rho = -3H$:
\be
{\dot M_s\over M_s}= (R_s^3\rho_b)^{-1}{d\over dt}(R_s^3\rho_b)= 3\left({\dot R_s\over R_s}-H\right) {\rm for }{\dot R_s\over R_s}> H; {\rm zero~otherwise.}
\ee
\be 
{d\over dt}{\dot R_s}= {8\pi p G \over \Omega_b H^2 R_s} - {3\over R_s}(\dot R_s -HR_s)^2 - (\Omega - {1\over 2} \Omega_b) {H^2 R_s\over 2}
\ee
\be
{\dot E}= L- p_bdV_s/dt = L - 4\pi p_b R_s^2 \dot R_s
\ee   
The physics expressed by these equations can be understood as follows. The mass of the shell increases in time as long as its velocity is larger than the Hubble expansion. The newly added material must be accelerated to the shell velocity, thus resulting in a net braking force. The internal pressure term has therefore to counteract both this force and the gravitational one. The third equation expresses energy conservation: the luminosity $L$ incorporates all sources of heating and cooling of the plasma. Typically these include the supernova energy injection, cooling by Compton drag against the CMB, bremsstrahlung and ionization losses. These equations have in general to be solved numerically. However, some of the main features of the blastwave evolution can be identified by a simple dimensional analysis. Three different regimes can be isolated during the evolution. At first, for bubble ages $t \ll t_H$ the gravity and Hubble flow are negligible and one can easily show that $R\propto t^{3/5}$ as in the more common case of non-cosmological blastwaves. When $t \approx t_H$, the behavior becomes quite complicated as several effects control the evolution at the same time: SN explosions have ceased, slowing the expansion; cooling and $pdV$ work reduces $p$ essentially to zero: the blast enters the momentum conserving phase in which $R\propto t^{1/4}$; 
gravity becomes important, decelerating the expansion. Finally, when the age becomes larger than the Hubble time $t \gg t_H$, the shell gets frozen into the Hubble flow, i.e. $R\propto t^{2/3}$ for $\Omega=1$. 

For our purposes here it will suffice to consider the size of the metal bubble in the Sedov-Taylor evolutionary phase, which can be obtained as the asymptotically-adiabatic, non cosmological limit of the previous equations. Then, we find
\be
V_Z(t) \equiv {4\pi \over 3 } R_s^3 = \zeta \left(\frac{{\cal E}}{\bar\rho}\right)^{3/5} t^{6/5}
\label{eq_ST}
\ee
where $V_Z$ is the metal bubble volume, ${\cal E}$ is the total energy released by the SNe driving the bubble, $\bar\rho$ is the mean IGM gas density, $t$ is the bubble age (say, the time elapsed from the explosion) and $\zeta$ is a dimensionless factor of order unity. In the simplest case, following the results of the previous Section, it is ${\cal E} = \nu E M_*$. Numerical work by Pallottini \etal (2014a) shows that all simulated bubbles expand for a similar maximum time, $t_{max} \approx 250$ Myr, essentially set by energy losses. So the final volume of the bubble is well approximated by eq. \ref{eq_ST} evaluated at $t = t_{max}$.

The analogous expression for the proper volume, $V_I$ of a cosmological \HII region can be derived from the following equation (Shapiro \& Giroux 1987):
\be
{dV_I \over dt } - 3 H V_I = f_{esc}{\dot{\cal N}_\gamma M_* \over \bar n_H} - {V_I \over t_{rec} } 
\label{eq_VI}
\ee 
where $f_{esc}$ accounts for the fraction of the ionizing photons escaping in the IGM.  This value is known to be of the order of a few percent in local and low-redshift galaxies, but it is conceivable that it could become $> 0.5$ in the small galaxies representing the dominant reionization sources (see discussion in Ferrara \& Loeb 2013). When the luminous source lifetime, $t_s$, is much shorter than $(t_{rec}, H^{-1})$, as it is often the case, recombinations can be neglected and the evolution of the \HII region can be decoupled from Hubble expansion. Thus, eq. \ref{eq_VI} has the simple solution 
\be
V_I \simeq f_{esc}{\dot{\cal N}_\gamma M_* \over \bar n_H} t_s = f_{esc}{N_\gamma M_*\over \bar \rho_H}
\ee 
Then, the relative ratio of the metal and ionization \textit{filling factor}, defined as the fraction of cosmic volume occupied by the bubbles, is
\be
{V_Z\over V_I} = \zeta {(\nu E t_{max}^2)^{3/5} \over f_{esc} N_\gamma}  \left({\bar \rho_H \over M_*}\right)^{2/5} \propto n_*^{2/5}
\ee 
where $n_*$ is approximately the mean comoving number density of galaxies of stellar mass $M_*$. From the previous expression we find that metal enrichment can more efficiently trail reionization (larger ${V_Z/V_I}$ ratio) if the dominant sources are more numerous or, equivalently, have lower masses. An early enrichment (often referred to as \textit{pre-enrichment} following Madau, Ferrara \& Rees 2001) has important general implications the we discuss next.

\subsection{Filling factor}  
First, the volume filling factor of enriched material, quantified by the porosity factor, $Q$, becomes large if pollutants are dwarf galaxies. This can be seen as follows. In a $\Lambda$CDM universe, structure formation is a hierarchical process in which nonlinear, massive structures grow via the merger of smaller initial units. Large numbers of low–mass galaxy halos are expected to form at early times in these popular cosmogonies, perhaps leading to an era of widespread pre–enrichment and preheating. The Press–Schechter (hereafter PS) theory for the evolving mass function of dark matter halos predicts a power–law
dependence, $dN/d\ln M_h \propto M_h^{(n_{eff}-3)/6}$. As hot, metal-enriched gas from SN-driven winds escapes its host halo, shocks the IGM, and eventually forms a blast wave, it sweeps a region of intergalactic space which increases with the 3/5 power of the energy, $E$, injected into the IGM (in the adiabatic Sedov–Taylor phase). The total fractional volume or porosity, $Q$, filled by these ‘metal bubbles’ per unit explosive energy density
$E dN/d\ln m$ is then 
\be
Q \propto E^{3/5} dN/d ln M_h \propto (dN/d \ln M_h)^{2/5} \propto M_h^{-41/150},
\ee
having assumed $n_{eff}=-2.2$ as stated above. Within this simple scenario it is the star–forming objects with the smallest masses which will arguably be the most efficient pollutant of the IGM on large scales. As a caveat we note that the above estimate ignores the clustering of galaxies as well as the possible re-accretion of ejecta. The latter effect, however, depends strongly on the modeling of the outflows and therefore remains quite uncertain. 

\subsection{Cooling time}  
The second point concerns the ability of the shocked gas to cool. This is necessary as
at $z = 3-4$, Ly$\alpha$ clouds show a spread of at most an order of magnitude in their
metallicity, and their narrow line widths require that they be photoionized and cold
rather than collisionally ionized and hot. At these redshifts, hot rarefied gas, exposed to
a metagalactic ionizing flux, will not be able to radiatively cool within a Hubble time. The
simple formula below, gives the redshift span required for a gas of primordial composition
(at the metallicities present in the Ly$\alpha$ forest contribution from metal cooling is virtually
negligible) at the mean cosmic density heated at some redshift $z_i$ to cool down:
\be
\Delta z = 231(\Omega_m h^2)^{1/2}(1 + z_i)^{-3/2} \approx 3\left({1+z_i \over 10}\right)^{-3/2} 
\ee
Hence, a gas that has been shock-heated at $z_i = 9$ will be already cooled by $z = 6$, but
if heating occurs at $z\approx 5$ the cooling time will exceed the Hubble time.

In conclusion early pre-enrichment by dwarfs offers the double advantage of a large
metal filling factor and efficient cooling of the metal enriched gas ejected by galactic
outflows. While it is possible that some metals are dispersed in intergalactic space at
late times, as hot pressurized bubbles of shocked wind and SN ejecta escaped the grasp of
massive galaxy halos and expanded, cooling adiabatically, into the surrounding medium,
such a delayed epoch of galactic super–winds would have severely perturbed the IGM
(since the kinetic energy of the ejecta is absorbed by intergalactic gas), raising it to
a higher adiabat and producing variations of the baryons relative to the dark matter.
Ly$\alpha$ forest clouds would not then be expected to closely reflect gravitationally induced
density fluctuations in the dark matter distribution, and the success of hydrodynamical
simulations in matching the overall observed properties of \Lya absorption systems would
have to be largely coincidental. In contrast, the observed narrow Doppler widths could
be explained if the ejection of heavy elements at velocities exceeding the small escape
speed of subgalactic systems were to take place at  high redshifts.

\section{Learning from simulations}  
A large number of studies based on cosmological simulations have attempted to study the evolution of the IGM metal enrichment (for a review see, e.g. Aguirre \etal 2007), both concentrating on the early pre-enrichment phases or making connection with observations which typically sample intermediate redshifts $3 < z <6$ in the post-reionization epoch. 

Early attempts to model the evolution of the cosmic metal distribution and the thermodynamic state of the polluted IGM date back to Springel \& Hernquist (2003a). Such model, in order to overcome the difficulties of producing substantial winds related to the extreme radiative energy losses in the high density gas surrounding the burst, developed a phenomenological wind model, in which the particles are assigned a given velocity and move ballistically for a certain amount of time. A similar strategy was also adopted by Bruscoli \etal 2003. Cen \etal (2005) took a different, simpler approach in which the energy and mass injection rates  are assumed proportional to the star formation rate. In their study they pointed out that winds from intermediate redshift galaxy populations (closely resembling Lyman Break Galaxies) contribute negligibly to the enrichment of the IGM, as these systems are relatively massive, thus forming too recently

In a series of papers Oppenheimer \& Dav\'e (2006, 2008, 2009) and collaborators (Oppenheimer \etal 2012) have studied in detail the IGM metal enrichment and proposed an alternative wind model, the so-called momentum-driven model in which the bubble expansion is driven by radiation pressure acting on an absorbing species (dust or free electrons). The resulting temperature of the bubbles is lower than in the standard shock-driven case; however, it is unclear if such a model performs better in explaining a number of observed properties of the bubbles as, for example, the cross-correlation of metal systems.

From all these numerical experiments, it was then realized that the huge dynamical range of the underlying physical phenomena makes a true self-consistent metal enrichment simulation nearly impossible at present. A viable modelization can be achieved by using subgrid models. These depend both on the considered physics and code implementation. Recently, Hopkins \etal (2013) studied the impact of different star formation criteria; Agertz \etal (2012) and Vogelsberger \etal (2013) analyzed the effects of different feedback prescriptions. The AQUILA project (Scannapieco \etal 2013) extended the previous studies to an analysis of 13 different feedback prescriptions used by popular cosmological codes. Subgrid modelling lessens the burden of the large dynamical range, but given the currently available computational capabilities the numerical resources have to be focused toward either the small or the large scales.

Simulations of small cosmic volumes, i.e. box sizes $\lsim\, 1 {\rm Mpc}\,h^{-1}$, concentrate the computational power and allow the usage of highly refined physical models. Tornatore \etal (2007) concentrated on the role of metal enrichment in inducing a transition from Pop III to Pop II stars (often dubbed as ``chemical feedback'') and it was the first study to include such process in a cosmological context. The key result is that Pop III star formation proceeds in a "inside-out" mode in which formation sites are progressively confined at the periphery of collapsed structures, where the low gas density and correspondingly long free-fall timescales result in a very inefficient star formation.
Greif \etal (2012) also studied the transition from Pop III to Pop II stars in a $10^{8}\msun$ galaxy at $z\sim10$ assessing the role of radiative feedback; Maio \etal (2010) analyzed the same transition by varying several parameters, such as the critical metallicity $\zcr$, the IMF, the metal yields and the star formation threshold; Xu \etal (2013) focalized on pinpointing the remnant of Pop III at high redshifts, by employing the same computational scheme of Wise \etal (2012), which analyzed the impact of radiation from first stars on metal enrichment at $z\gsim9$; at the same redshift, Biffi \& Maio (2013), using an extensive chemical network, studied the properties and the formation of first proto-galaxies.

Larger scale ($\gsim 10\, {\rm Mpc}\,h^{-1}$) cosmological simulations are better suited for a fair comparison with the observations. Scannapieco \etal (2006) showed that observation of line of sight (l.o.s.) correlations of $\CIV$ and $\SiIV$ are consistent with a patchy IGM enrichment, confined in metal bubbles of $\sim 2 \, {\rm Mpc}\,h^{-1}$ at $1.5\lsim z\lsim3$; by implementing galaxy outflows driven by a wind model Oppenheimer \& Dav\'e (2006) managed to reproduce the flatness of $\Omega_{\rm CIV}$ at $2\lsim z\lsim5$. Cen \& Chisari (2011) simulated a 50 Mpc~$h^{-1}$ box, and find, among other results, a good agreement with observations for the $\Omega_{\rm CIV}$ evolution and a reasonable match for $\Omega_{\rm OVI}$.  By using a $(37.5\,{\rm Mpc}\,h^{-1})^{3}$ volume simulation evolved up to $z=1.5$, and considering different IMFs and feedback mechanisms, Tescari \etal (2011) analyzed the evolution of $\Omega_{\rm CIV}$ and statistics of \HI\, and $\CIV$ absorbers at different redshifts. With a box size of $25 \, {\rm Mpc}\,h^{-1}$ and including various feedbacks, Vogelsberger \etal (2013) managed to match several observations, such as the SFR and stellar mass density (SMD) evolution for $z\lsim 9$, the galaxy stellar mass function and mass-metallicity relation at $z=0$. Finally, Pallottini \etal (2014a) presented an extensive AMR hydrodynamical simulations in a $(10 {\rm Mpc}\,h^{-1})^3$ volume up to $z = 4$, including the Pop III - Pop II transition, and following the joint evolution of metal enrichment on galactic and intergalactic scales.
\begin{figure*}
\centering
\includegraphics[width=8.3cm]{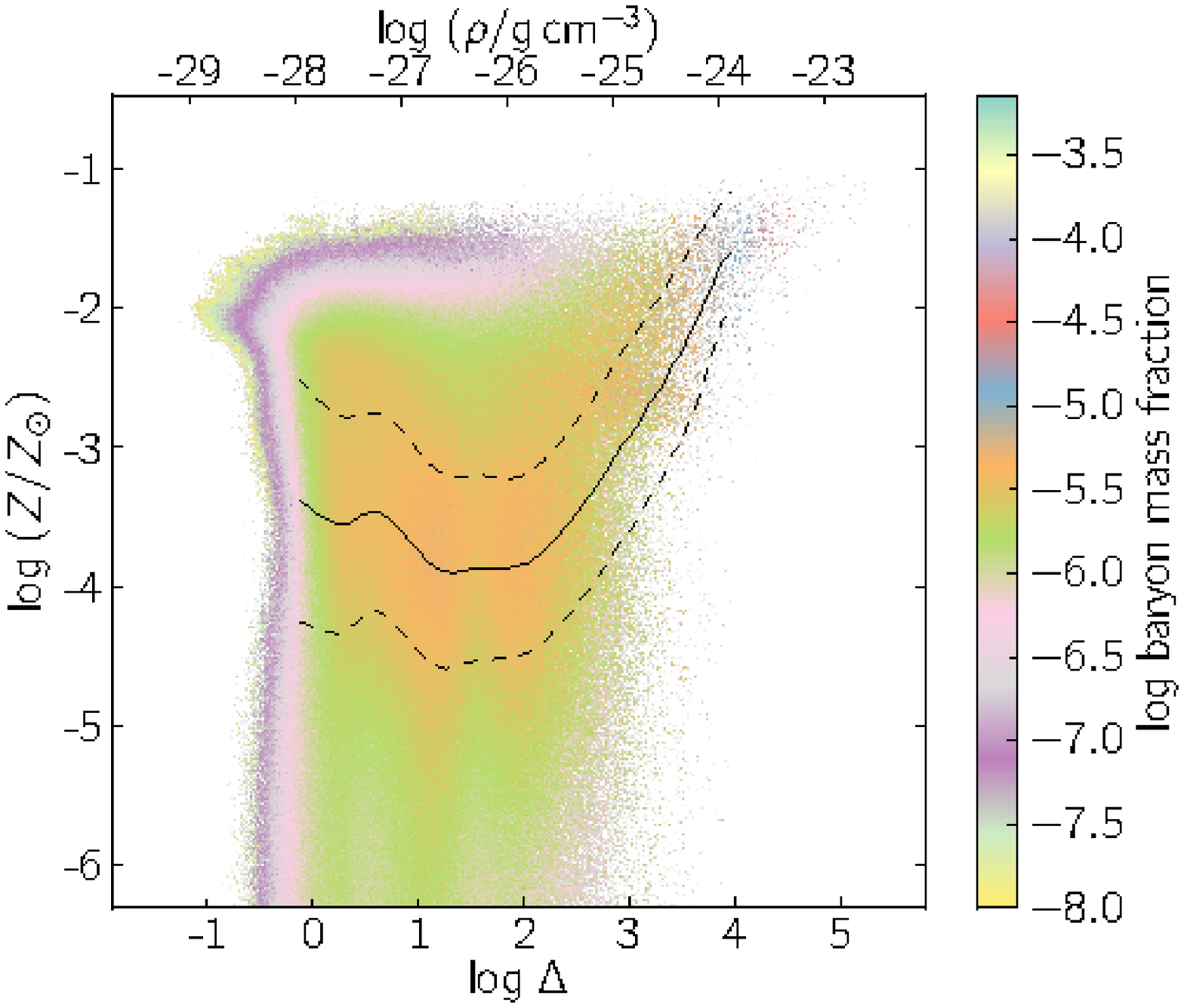}
\includegraphics[width=8.3cm]{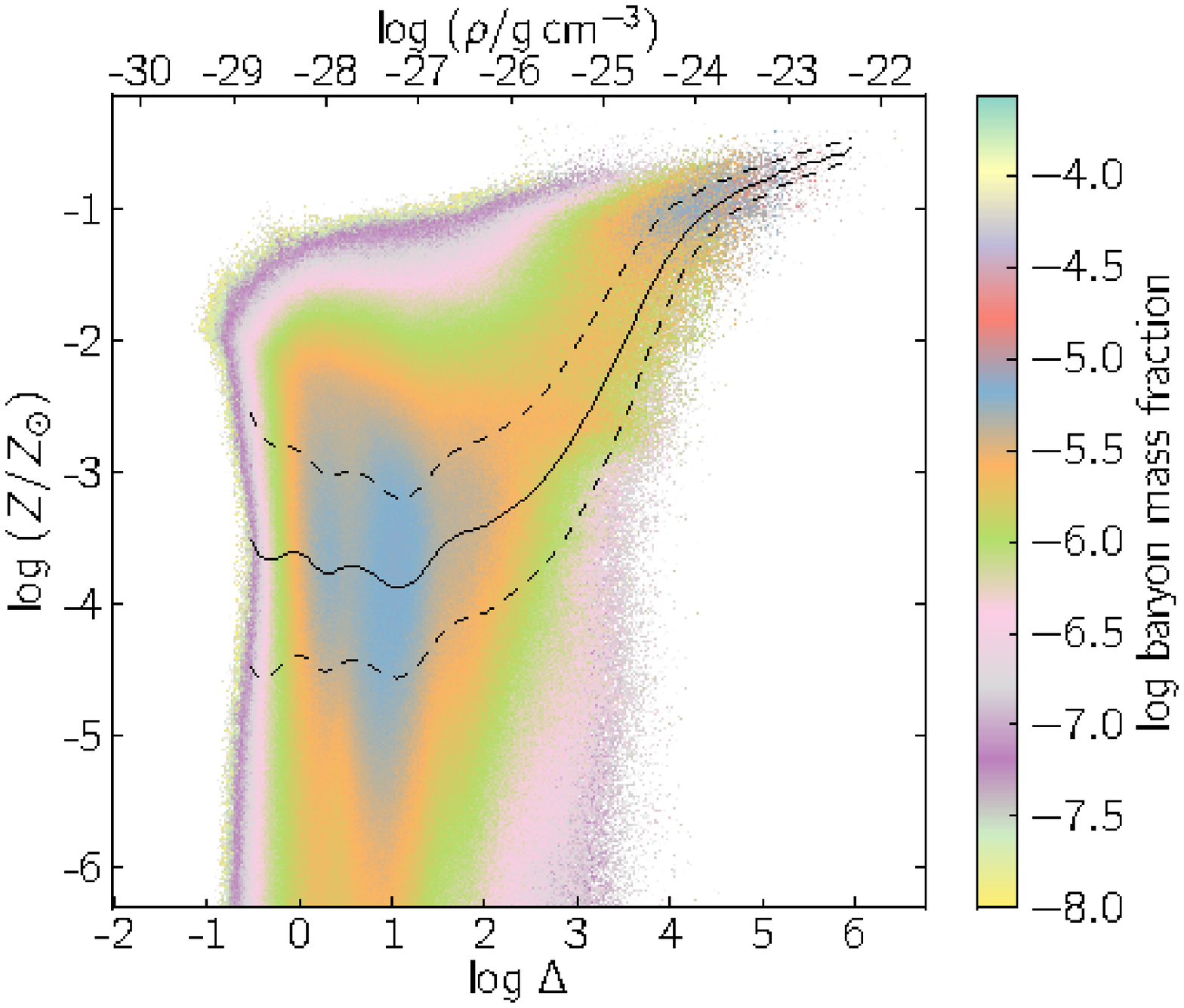}
\caption{Mass weighted probability distribution function (PDF) of the baryons at $z=6$ (top) and $z=4$ (bottom) in the metallicity-overdensity plane. The solid (dashed) black line is the mean (r.m.s.) metallicity as a function of density. Taken from Pallottini \etal (2014a). \label{fig_z_stato_04}}
\end{figure*}
Clearly, it is impossible to single out the effects of metal enrichment, galaxy evolution and cosmic reionization. On the contrary, there is a strong inter-relationship among them which adds further complexity to the problem. For example, cosmic reionization might suppress the star formation within galaxies, thus decreasing the metal production. On the other hand, an increased metal content of the gas 
enhances the cooling and hence the ability of galaxies to ultimately form stars. These and many other relevant processes are very difficult to be properly included in the simulations and therefore the emerging scenario is necessarily sketchy. A few points seem, however, relatively solid and agreed upon by studies based on  state-of-the-art simulations (Oppenheimer, Dav\'e \& Finlator 2009; Tescari \etal 2009; Cen \& Chisari 2011; Barai \etal 2013; Pallottini \etal 2014a).

At $z\approx 5-6$ galaxies account for $\lsim 10\%$ of the baryonic mass; the remaining gas resides in the diffuse phases: (a)~\textit{voids}, i.e. regions with extremely low density ($\Delta\leq 1$), (b) the true \textit{intergalactic medium} (IGM, $1<\Delta\leq 10$) and (c) the \textit{circumgalactic medium} (CGM, $10<\Delta\leq 10^{2.5}$), the interface between the IGM and galaxies. Contrary to baryons, which reside predominantly in the IGM, metals are found at any given redshift primarily near their production sites, i.e. in galaxies. However, while at $z=6$ metals in the ISM make up to about 90\% of the total heavy elements mass, at later epochs ($z=4$) this fraction increases to 95\% as a result of the increased ability of collapsed objects to retain their nucleosynthetic products thanks to their larger potential wells. Among the diffuse components, at $z=6$ the CGM is more enriched than the IGM (voids) by a factor 1.6 (8.8), as metals cannot be efficiently transported by winds into distant, low-density regions. Interestingly though, even the most diffuse gas in the voids has been polluted to some extent. 
\begin{figure}
\centering
\includegraphics[width=8.3cm]{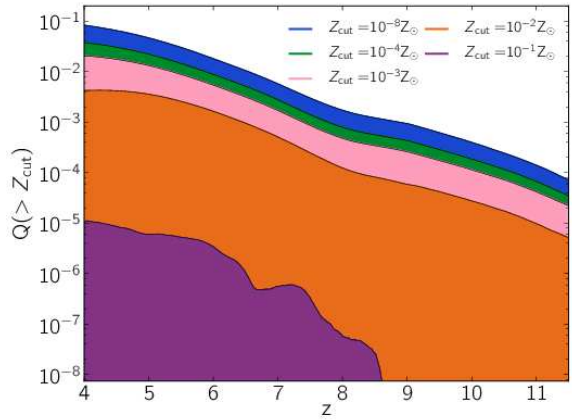}
\caption{Redshift evolution of the volume filling factor, $Q(>Z_{cut})$, of regions enriched to a metallicity $Z >Z_{cut}$. Taken from Pallottini \etal (2014a). \label{fig_z_stato_04}
\label{filling}
} 
\end{figure}

The $Z-\Delta$ distribution of the cosmic gas provides additional insights in the metal enrichment process (Fig. \ref{fig_z_stato_04}). At $z=6$ (top panel) baryons are nearly uniformly distributed in $10^{-1}\lsim\Delta\lsim 10^{2.5}$ and the cosmic gas is characterized by a broad range of metallicities ($10^{-6} \lsim {Z}\slash Z_\odot \lsim 10^{-2}$). Besides containing most of the metals, galaxies ($\Delta\gsim10^{2.5}$) show high metallicities ($10^{-2} \lsim Z\slash Z_\odot \lsim 10^{-1}$) and a loose $Z-\Delta$ correlation. At $z=4$ (bottom panel) the distribution evolves and the $Z-\Delta$ correlation at high density becomes tighter and steeper, additionally extending to lower overdensities. Both the IGM and the CGM become preferentially enriched at $10^{-4.5}\lsim Z\slash Z_\odot\lsim10^{-2.5}$, i.e. around the critical metallicity for the Pop III transition. This is consistent with the simple analytical estimates above. However the contribution of Pop III stars to metal enrichment is negligible, as expected from analytical arguments given above. 

As bubbles age, metals are not only produced at an increasing rate but they are transported by winds away from the production sites. To see this it is useful to look at the fraction of the cosmic volume, $V$, filled with heavy elements at a metallicity larger than a given value, $Z_{cut}$ . The behavior of $Q$ for different values of $Z_{cut}$ has been traced on-the-fly in the simulation. The result is shown in Fig. \ref{filling} (from Pallottini \etal 2014a), which highlights interesting features of the enrichment process. As a reference we note that the filling factor of the regions with $Z > 10^{-3} Z_\odot$ at $z=6$ is about 1\%.

\begin{figure}
\centering
\includegraphics[width=12.3cm]{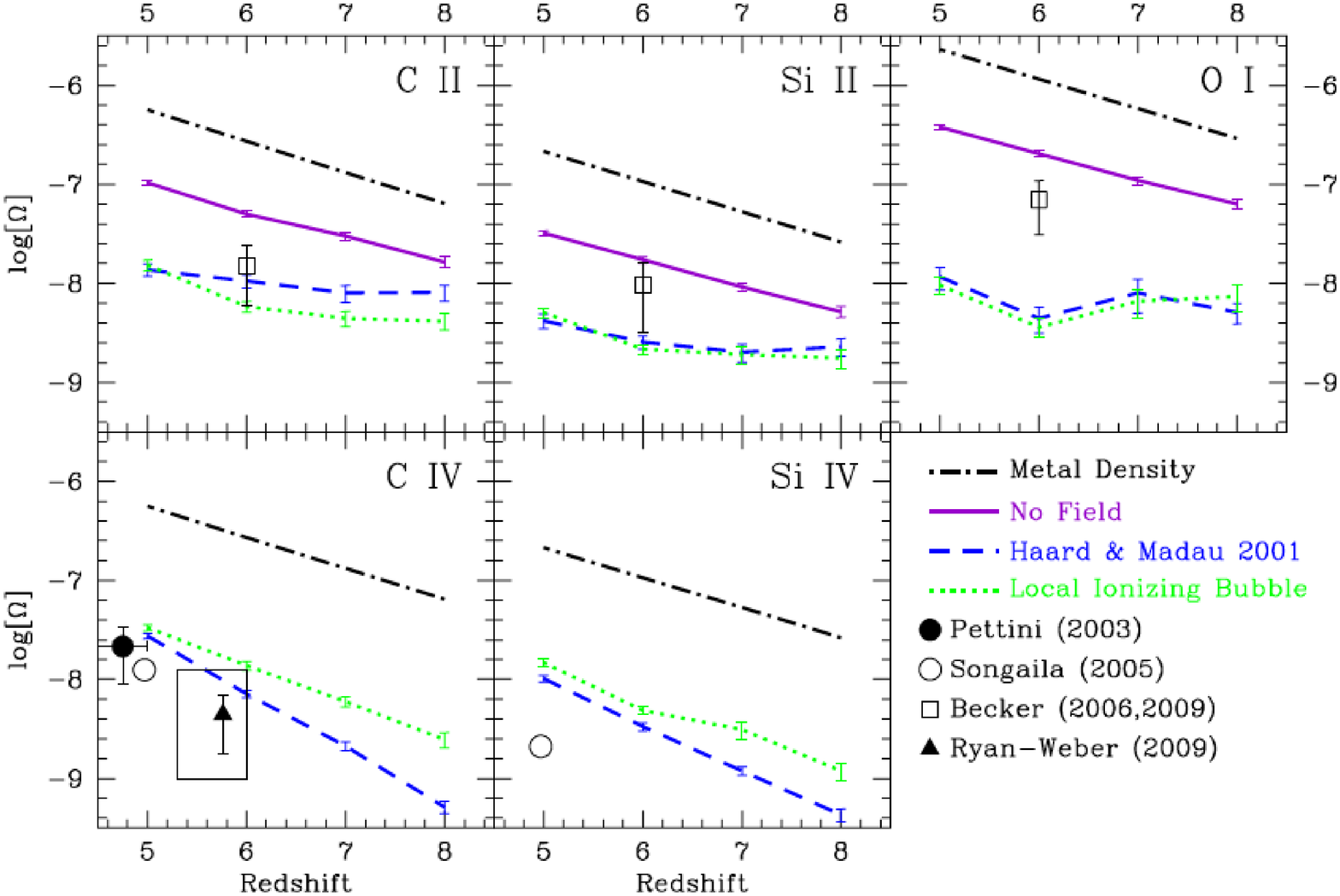}
\caption{Redshift evolution of various species for the three ionization cases compared to the total elemental abundance. The No Field case assumes no ionization correction. Most \CIV observations cannot distinguish between ionization produced by a background (Haardt \& Madau)  or a local source, i.e. a nearby galaxy (Bubble), although the rapid evolution observed by Becker et al. (2009) supports the former. Taken from Oppenheimer, Dav\'e \& Finlator (2009)
\label{evol}
} 
\end{figure}

Detectable absorbers generally trace inhomogeneously-distributed metals residing outside of galactic halos. \CIV is an ideal
tracer of IGM metals at $z \approx 6$;  simulations show a strongly increasing global $\Omega_{\rm CIV}$ in $z = 8 \rightarrow 5$, in contrast to its relative constancy from $z = 5 \rightarrow 2$. It is unlikely that the drop is due to a decrease in the actual abundance of C, but most likely it is the result of a weaker/softer UV background, favoring lower ionization states. Fig. \ref{evol} taken from Oppenheimer, Dav\'e \& Finlator (2009) clearly shows this point for \CIV and other species.

\section{Learning from observations}
Historically, metal enrichment of the intergalactic medium has been studied by analyzing the absorption features imprinted by intervening material along the line of sight in the spectra of distant luminous sources, mostly quasars. These experiments allow us to precisely infer key quantities as the column density of the absorbing species, the gas ionization state, temperature, and peculiar velocities in a statistically significant manner. It is also possible to study the column density and redshift distribution of metals and the clustering properties of the absorbing systems. The bottleneck of this technique consists in the availability of sufficiently bright sources at high redshifts from which high signal-to-noise spectra can be obtained. The rapid decline of luminous quasars beyond $z>6$ in this sense represents a problem. In principle a similar use of Gamma-Ray Bursts can be also conceived (Totani \etal 2006; Gallerani \etal 2008; Wang \etal 2013) to study both metal enrichment and reionization but unfortunately so far it has been proven difficult to obtain sufficiently high-quality spectra enabling a thorough analysis. It is even more difficult to use the comparably fainter galaxies as background sources.

The interpretation of the metal line data provides a strong test of this metal enrichment. Until recently, most studies have concentrated on highly ionized species, as  OVI(1032, 1038), NV(1239, 1243), CIV(1548, 1551), SiIV(1394, 1403), and CII(1335), where numbers indicate the wavelength in Angstroms.  With the exception of the OVI doublet they all have rest wavelengths $\lambda_0 > 1216$ \AA, implying that for sufficiently high redshifts, they appear redward of the Ly$\alpha$ forest and hence are relatively easy to detect. As the ionization state of these metals is governed by the intensity and spectrum of the ionizing cosmic UV background, such high ionization states are typically found in the post-reionization epoch or in the vicinity of a starburst galaxy. 

In the last few years, following the initial suggestion by Oh (2002), and with the aim of studying metal enrichment at redshifts well into the reionization epoch, the interest has somewhat shifted to low ionization species as CII (1334), OII (1302), Si II (1260). These species are typically associated with more neutral gas and therefore are better suited to trace the \HI pockets present prior to reionization completion. 

The key results and present understanding of the field can be summarized as follows. The analysis of a considerable number ($\approx 100$) of \CIV systems  from high quality X-Shooter spectra at $z=6$ has shown that the distribution of this species evolves considerably from $z \approx 3$. The evolution is predominantly driven by the stronger lines in the column density range $13.8 < \log N_{\rm CIV} < 15$, while the evolution is much weaker (or almost absent) in the low $N_{\rm CIV} $ systems. The density evolution of the \CIV abundance is well fitted by (D'Odorico \etal 2013) a power law
\be
\Omega_{\rm CIV} = (2\pm 1) \times 10^{-8} \left({1+z\over 4}\right)^{-3.1 \pm 0.1}     
\ee
Consistent results, albeit at a lower statistical significance, are obtained from the analysis \SiIV absorption lines; both species trace well the IGM at low-to-moderate overdensities ($\Delta \approx 10$). The decrease of the \CIV abundance was already noted by Becker \etal (2009), who, however, were only able to put an upper limit on $\Omega_{\rm CIV} < (0.4-1) \times 10^{-8}$ (the uncertainty is related to the unknown slope of the columns density distribution ($f(N_{\rm CIV}) \propto N_{\rm CIV}^{-\alpha}$) in $5.3 < z < 6$). Thus, this value is basically consistent with the more recent result above. The rapid evolution of $\CIV$ at these redshifts suggests that the decrease in the number density may largely be due to ionization effects, in which case many of the metals in the $z=4-5$ IGM could already be in place at $z > 5.3$, but in a lower ionization state. 

Clearly, because of the unknown ionization correction factor (necessary to translate the observed \CIV into a C column density), and elemental abundance pattern, obtaining an IGM metallicity value to be readily compared with the theoretical predictions discussed above is challenging. To overcome the problem simulations (see below) are often used in combination with photoionization codes, such as CLOUDY, to predict the abundance of observed species, under a given evolutionary model for the ionizing radiation filed, which -- in general -- can be a combination of the UV background and local effects from galaxies.  As a rule of thumb, though, if \CIV is the dominant ionization stage with ionization fraction of 0.5, and assuming a solar abundance for carbon of  $3.3 \times  10^{-4}$, one gets that  $\Omega_{\rm CIV} = 2 \times 10^{-8}$ can be translated in a metallicity of $\approx 10^{-4} Z_\odot$, very consistent with what was deduced in eq. \ref{key}.

Peering into the reionization epoch requires us to turn our attention to low ionization species.  These experiments become critical given the observed decline of \CIV towards high redshifts. As we have seen, reionization requirements in terms of ionizing photons imply that an amount of metals corresponding to $\approx 10^{-4} Z_\odot$ must be produced before the end of the process, i.e. by $z\approx 6$. Then, if these metals are not in high ionization species, they must be found in low ionization states. \OI offers the most promising way to clarify this issue. The detection of a ``\OI forest'' would suggest that (a) the
IGM is at least partially metal-enriched by $z>6$, and (b) that reionization completed relatively late with respect to expectations based on CMB studies. So where do we stand? 

The most comprehensive searches for \OI absorption at high redshift have been performed by Becker \etal (2007) and Becker \etal (2011b). The latter survey uses spectra of 17 QSOs in the redshift range $5.3 < z < 6.4$ and detected 10 low-ionization metal systems; 9 of these systems contain \OI lines; however, none of them shows the strong \CIV or \SiIV that are commonly found in lower redshift analog (sub-Damped Lyman Alpha) systems.   An analysis performed by Keating \etal (2014) suggests that these absorbers have densities $\Delta \approx 80$ and metallicities $2\times 10^{-3} Z_\odot$. These overdensities are large enough that they correspond to the so-called Circum-Galactic Medium (CGM, see above), \ie an interface between a galaxy and the surrounding IGM. 

It then appears that the enrichment around low-mass galaxies has already progressed considerably by $z \approx 6$. Interestingly, the decrease in the photoionization rate causing the drop in the \CIV, boosts the incidence rate of \OI systems, i.e. the number of absorbers per line of sight, towards high redshift.  A similar effect is expected for \CII, a species that should also be favored by a lower IGM ionization state. 
This is particularly exciting because, as deduced from their large number, low-ionization absorption lines might be already probing the environment of very faint galaxies, possibly responsible for reionization, that cannot be detected in $i$-dropout and Ly$\alpha$ emission galaxy surveys. 

Thus, as the volume filling factor of the metals is becoming smaller towards higher redshift (Fig. \ref{filling}) we expect that absorption features become more concentrated around galaxies. As a result, high redshift absorption line experiments probe the CGM of the most numerous - and hence smaller - galaxies. However, the proximity of observed metals to their parent galaxies must be reconciled with the disappearance of higher excitation lines, which should in principle be preserved by the increased UV radiation coming from the galaxy.  These studies will then allow to understand in detail the interplay between galaxy formation, winds and reionization in an unprecedented manner. As this exciting perspective has become possible only recently, to gain experience and get some guidance, a number of recent works have appropriately concentrated on the study of lower redshift systems, where experiments are less challenging.

The CGM has been probed so far up to $z \approx 3$ using absorption lines of both H I (e.g. Rudie et al. 2012, 2013; Pieri et al. 2013) and heavy elements (e.g. Steidel et al. 2010; Churchill et al. 2013; Borthakur et al. 2013; Jia Liang \& Chen 2014). These observations show that the CGM extends up to impact parameters $b \simeq 10 r_{vir}$, where $r_{vir}$ is the virial radius of the parent dark matter halo. An anti-correlation between the absorption equivalent width, EW, and $b$ is observed. Moreover, the EW profiles appear to be self-similar once scaled with $r_{vir}$, with no signs of evolution from $z=2$ to $z=0$ (Chen (2012) in agreement with the findings from hydrodynamical simulation by Pallottini, Gallerani \& Ferrara (2014b), who also were able to explain the EW-$b$ anti-correlation in their model. Thus, if the properties of the CGM are not evolving with redshift, understanding the physics of the high redshift systems should represent less of a challenge.

\vskip 1truecm
I am indebted to A. Pallottini, S. Gallerani, L. Vallini and B. Yue for discussions and collaboration on aspects relevant to the subject of this review.

\end{document}